\documentclass[useAMS,usenatbib]{mn2e}

\usepackage{graphicx}
\usepackage{amsmath}
\usepackage{amssymb}

\title[Optical filaments in galaxy clusters]{The generation of optical
emission-line filaments in galaxy clusters}

\author[E.C.D. Pope, J.M. Pittard, T.W. Hartquist, S.A.E.G. Falle]
{Edward C.D. Pope$^{1}$\thanks{E-mail:e.c.d.pope@leeds.ac.uk}, Julian
M. Pittard$^{1}$, Thomas W. Hartquist$^{1}$, Sam A.E.G. Falle$^{2}$\\
$^{1}$School of Physics \& Astronomy, University of Leeds, Leeds, UK,
LS2 9JT\\$^{2}$School of Applied Mathematics, University of Leeds,
Leeds, UK, LS2 9JT\\ }

\begin{document}

\pagerange{\pageref{firstpage}--\pageref{lastpage} \pubyear{2007}}

 \maketitle

\label{firstpage}

\begin{abstract}

\noindent Recent data support the idea that the filaments observed in
H$\alpha$ emission near the centres of some galaxy clusters were
shaped by bulk flows within their intracluster media. We present
numerical simulations of evaporated clump material interacting with
impinging winds to investigate this possibility. In each simulation, a
clump falls due to gravity while the drag of a wind retards the fall
of evaporated material leading to elongation of the tail. However, we
find that long filaments can only form if the outflowing wind velocity
is sufficiently large, $\sim 10^{8}\,{\rm cm\,s^{-1}}$. Otherwise, the
tail material sinks almost as quickly as the cloud. For reasonable
values of parameters, the morphological structure of a tail is
qualitatively similar to those observed in clusters. Under certain
conditions, the kinematics of the tail resemble those reported in
Hatch et al.(2006). A comparison of the observations with the
numerical results indicates that the filaments are likely to be a few
tens of Myrs old. We also present arguments which suggest that the
momentum transfer, from an outflowing wind, in the formation of these
filaments is probably significant. As a result, tail formation could
play a role in dissipating some of the energy injected by a central
AGN close to the cluster centre where it is needed most. The trapping
of energy by the cold gas may provide an additional feedback mechanism
that helps to regulate the heating of the central regions of galaxy
clusters and couple the AGN to the ICM.
\end{abstract}


\section{Introduction}

Optical emission-line nebulae commonly surround massive galaxies in
the centres of X-ray bright, cool cluster cores \citep[][]{crawf}. The
origin of the H$\alpha$ filaments has been attributed to a variety of
processes including: condensation from an intracluster medium (ICM)
evolving as a cooling flow \citep[][]{fab84, heckman, don91};
accretion of clouds captured in galaxy mergers \citep[][]{braine};
expulsion from the central galaxy \citep[][]{burb}.

NGC 1275 at the centre of the Perseus cluster contains the best
studied example of such a nebula \citep[][]{hatch}. Its filaments are
typically 50-100 pc thick and up to 30 kpc long, and the majority of
them are radial. \cite{hatch} presented kinematic data that rules out
dynamical models of purely infalling filaments. The observed kinematic
properties provide strong evidence that the filaments are not in
gravitational free fall, because, if they were, their velocities would
rise sharply towards the centre of the nebula \citep[][]{heckman}. The
most conclusive evidence lies in the velocity structures of the
northern and northwestern filaments. The lower half of the northern
filament is redshifted with respect to the galaxy, whilst the upper
section is blueshifted. Thus, the upper part of the filament is
flowing away from the galaxy while the lower part is flowing into the
galaxy.

The filaments may follow streamlines of the flow of more tenuous
material around them. In the Perseus cluster, some filaments appear to
have been shaped by the wakes of the buoyant bubbles. Given that the
data suggest that, at least, some parts of the filaments are
outflowing, their origin may lie within the galaxy. NGC 1275 contains
a large reservoir of cold molecular gas that could fuel them.

The filaments have morphologies similar to those of laminar jets, and
their relatively smooth structures may place constraints on turbulent
motions in the ICM. Alternatively, an ordered, amplified magnetic
field trailing behind a buoyant bubble interacting with filaments may
prevent the destruction of the filaments by turbulent motions
\citep[][]{rusz07}.

The optical nebula around NGC 1275 emits $4.1 \times 10^{42}{\rm
erg\,s^{-1}}$ in H$\alpha$ and N[II]. The nature of the power source
remains unclear. Various excitation mechanisms for these lines have
been proposed. For instance, although ionisation by the central AGN
may be important for the inner regions, it cannot be the dominant
source of power for the extended nebula because the H$\alpha$
intensity does not decrease with distance from the nucleus
\citep[][]{john88}. The nebula may be excited by stellar UV, but there
is no spatial correlation between the filaments and the stellar
clusters. Excitation by X-rays from the ICM seems unlikely, as the ICM
may be as much as a hundred times less luminous in UV than in the
X-rays \citep[][]{fab03b}. Heat conduction from the ICM to the colder
filaments has also been proposed \citep[][]{don00}, but it might also
lead to the evaporation of the filaments on too short a timescale
\citep[][]{nb04}. Shocks and turbulent mixing layers might play a role
\citep[][]{crawf92}. Cosmic rays, preferentially diffusing along the
magnetic field lines trailing behind rising bubbles, could possibly
drive the excitation in those filaments that are located in bubble
wakes \citep[][]{rusz07}. The same magnetic fields lines could also
prevent the filaments from evaporating due to thermal conduction.

The same mechanisms that power the emission could also have a profound
effect on the morphology of the emitting region. It is impossible to
take all of these processes into account. Consequently, we will
investigate the effect of gravity and an impinging wind that strips
material from a cloud on the morphology and kinematics of the
resulting filament.

The main aim of the work reported here is the calculation of the
density and velocity structures of material evaporated from clumps
embedded in winds from the central galaxies of galaxy clusters and the
comparison of model results with observations. Section 2 contains some
preliminary considerations, while the model assumptions and numerical
approach are treated in section 3.  The simulation results appear in
section 4. In section 5 we investigate the development of tails in
realistic environments by including the appropriate density and
gravity for a galaxy cluster. Section 6 concerns momentum tranfer
between clumps and the winds surrounding them. Section 7 discusses the
possible trends in optical emission between clusters, and section 8 is
a summary.

\section{Preliminary considerations}
\subsection{Filamentary widths}

We consider a clump, or cloud, embedded in a lower-density, hotter
ambient wind with a speed $v_{\rm w}$ in the frame of the central
galaxy. Material evaporates from the clump and interacts with the wind
to form a filament.

It may be possible to estimate the age of an observed filament from
its width. Previous numerical studies have shown that for a broad
range of parameters the width of a filament is comparable to the
diameter of the clump from which material evaporates
\citep[e.g.][]{dyson}. The radius of the filament $r_{\rm f}$ would
have a lower bound of approximately: $r_{\rm f} \sim 0.5 c_{\rm
s,c}\Delta t$, where $\Delta t$ is the age of the filament, and
$c_{\rm s,c}$ is the sound speed in the clump at the head of the
filament. The length of the filament $l_{\rm f}$ would have a maximum
given roughly by $l_{\rm f} \sim 2 r_{\rm f}v_{\rm w}/c_{\rm s,c}$.
The largest observed $l_{\rm f}/r_{\rm f}$ ratios would require that
$v_{\rm w}/c_{\rm s,c} \sim 100$s which is plausible if the clump
material is molecular (and, thus, cold) before being heated to the
filament temperature.

Alternatively, the width of the filament may be governed by the
diffusion of material into the ICM. Consequently, it is possible to
set an upper bound to the diffusion coefficient. The root mean square
(RMS) displacement of a particle after
time $t$ is $\sqrt{\langle r^{2} \rangle} = \sqrt{6 Dt}$, where $D$ is
the diffusion coefficient. If the RMS displacement due to diffusion
equals the measured radius of a filament, the filament's length is
approximately
\begin{equation}
l_{\rm f} \sim v_{\rm w}\Delta t \sim v_{\rm w} \frac{r_{\rm f}^{2}}{6
D}.
\end{equation}
Substituting order of magnitude estimates: $l_{\rm f} = 30$ kpc,
$r_{\rm f}= 100$ pc, $v_{\rm w} = 10^{3}{\rm km\,s^{-1}}$ we find that
$D \sim 10^{25}{\rm cm^{2}\,s^{-1}}$. Interestingly, this is
comparable to the diffusion coefficients found by \cite{roed} at 30 kpc in
their simulations of metal diffusion in the Perseus cluster. If
diffusion were contributing significantly to filament broadening, this
value of the diffusion coefficient would require the filaments to be
roughly $\Delta t \sim 5\times 10^{7}$ yrs old.

This diffusion coefficient greatly exceeds the Spitzer thermal and
viscous diffusivities. Turbulent diffusion would have to be important
for such a large diffusion coefficient to obtain. The turbulent
diffusion coefficient is given approximately by $D \sim vl/3$ where
$v$ is the characteristic velocity and $l$ the characteristic length
scale of the largest eddies.  If the turbulence is driven by the shear
of the fluid flow past the clump and along the filament, we might
assume that the thickness of the boundary layer is roughly 3 \% of the
length of the filament \citep[][]{hart,canto} which exceeds the
observed filament widths. Therefore, a diffusion picture is not
appropriate for the dynamics of the filament, but clearly viscous
coupling across the entire cross section of a filament is strong,
leading to a flow velocity that depends on the distance along the
filament but not on the location relative to the nearest edge.

\subsection{Clump mass-loss}

The rate at which mass is ablated from a clump, $\dot{m}$, might be
estimated by the rate at which momentum associated with the wind is
transferred to the clump, divided by a speed, $c_{\rm s,a}$, anywhere
between the sound speed of material in the clump and the difference
between the average clump speed and the average speed of material in
the filament. The appropriate value of $c_{\rm s,a}$ depends on how
effective the viscous coupling (due to turbulence or any other
mechanism) between the filament material and the wind material is. The
total mass lost by a clump would then be given by
\begin{equation}
m \approx \dot{m}\Delta t = \Delta t \frac{A\rho_{\rm w} v_{\rm
w}^{2}}{c_{\rm s,a}}.
\end{equation}
where A is the clump cross section and $\rho_{\rm w}$ is the mass
density of the wind.  If $\Delta t$ is 30 Myr, the clump has a radius
$r_{\rm c} \sim 50$ pc, $\rho_{\rm w} = 10^{-26} {\rm g\,\,cm^{-3}}$,
$v_{\rm w} = 10^{3} {\rm km\,\,s^{-1}}$ and $c_{\rm s,a} = 0.3 \rm
km\,\,s^{-1}$, the above equation gives $m\sim 10^{8} M_{\odot}$. This
is comparable to the mass of a structure with a number density of 10
$\rm cm^{-3}$, radius of 50 pc, and length of 30 kpc.

\subsection{Clump motion}

A model of the clump's motion taking into account the loss of mass and
the ram pressure of the impinging wind gives,
\begin{equation}\label{eq:mdot2}
\frac{{\rm d}}{{\rm d}t} (m\dot{x}) = -\dot{m}\dot{x} -mg + KA\rho_{\rm
w}(v_{\rm w}-\dot{x})^{2}.
\end{equation}
$K$ is a multiplicative factor which in the simplest case would be a
constant of order unity. $g$ is the gravitational field strength, $x$
is the height of the clump above the central galaxy's mid-plane, and a
dot indicates differentiation with respect to time.  Expansion of the
left-hand-side of equation (\ref{eq:mdot2}) demonstrates that mass
loss from the clump does not affect the motion of the
clump. Therefore, the terminal velocity of the clump is,
\begin{equation}\label{eq:term1}
\dot{x} = v_{\rm w} - \bigg(\frac{mg}{KA\rho_{\rm w}}\bigg)^{1/2}.
\end{equation}
Using the values given above for $\rho_{\rm w}$ and $r_{\rm c}$ and
taking $g\sim 10^{-8} {\rm cm\,\, s^{-2}}$, $K = 1$, and $m\sim 10^{5}
M_{\odot}$, we find that $(mg/KA\rho_{\rm w})^{1/2} \sim 10^{3} {\rm
km\,s^{-1}}$, and so is significant for slower outflow velocities. For
larger masses, and gravities, the terminal velocity is greater. The
terminal velocity of the clump is not likely to be constant, because
the mass of the clump falls with time. Alternatively, if the drag term
is negligible, then the clump will accelerate freely, and the velocity
will be given by $\dot{x} = gt$, if $g$ is a constant. However, this
case appears to be ruled out by observations showing slower
velocities.

Since material is continually stripped from the clump as it moves
relative to the surroundings, it may eventually disappear
completely. Clearly, if this happens during a timescale that is less
than that required for the velocity to exceed the terminal velocity,
then we need not simulate the case of a clump moving at its terminal
velocity. Assuming constant gravity, the free-fall time from 30 kpc
is about $10^{8}$ yr for the value of $g$ assumed above. If the clump
started from that height with zero velocity with respect to the centre
of the galaxy, it will have reached a speed of
roughly 400 km $\rm s^{-1}$ in the free-fall time.
Depending on the wind parameters and gravitational field,
a $10^8 M_{\odot}$ molecular clump will not reach terminal velocity and
may not be totally evaporated away during free fall from 30 kpc.

\subsection{Origin of the clouds}

As mention above, the data suggest that much of the material within
the filaments is outflowing. There are two scenarios in which
filaments with the observed properties could be produced. Firstly,
clouds may form in cooling gas accreting from outside the galaxy. As
they fall inwardly they would be stripped by an outflowing wind, and
the lower regions of the filament could be infalling while the outer
portion could be outflowing. Alternatively, such clouds may condense
in gas that has been expelled from the central galaxy, and eventually
begin falling back into the central galaxy as in a galactic fountain
\citep[e.g.][]{shap}. As the clouds fall inwardly, they would be
stripped by outflowing material. In either case, an outflowing wind
would ensure both infalling and outflowing portions of a filament.

\subsection{Outflows}

In a previous subsection we discussed the idea of ram pressure due to
an impinging wind, or outflow. However, it is unclear how precisely
how fast these winds are: the density and emission structures of
transonic flows can be similar to those of nearly hydrostatic
media. So, in the absence of X-ray observations with high spectral
resolution, direct observational diagnosis of the kinematics is unable
to distinguish firmly between low Mach number and transonic
speeds. Despite this, X-ray maps do show evidence of weak shocks in
the hot material at radii in the range of a few kpc to at least a few
10's of kpc in the Virgo and Perseus clusters
\citep[e.g.][]{form8707,sanders07}. Thus, we know that there are
large-scale flows with transonic speeds. The sound speeds of $3 \times
10^{7}$K and $10^{8}$K ionised gas, for the Virgo and Perseus
clusters, are $8 \times 10^{7}\,{\rm cm\,s^{-1}}$ and $1.5 \times
10^{8}\,{\rm cm\,s^{-1}}$, respectively. In conjunction with the X-ray
maps, this suggests that the wind speeds are likely to be $\sim
10^{8}\,{\rm cm\,s^{-1}}$. In fact, we suspect that the flows may be
fountain-like and that outwardly flowing gas eventually falls back
inwards. As a result, the wind velocity is not constant, but drops off
with radius. So by the time it reaches the cold clouds the wind
velocity will be much reduced from its central value.

AGN activity is considered a prime candidate for preventing the ICM
cooling from $10^{8}$K at rates of more than $100\, {\rm
M_\odot\,yr^{-1}}$. This requires an energy input rate of
$10^{44}\,{\rm erg\,s^{-1}}$ which is comparable to the kinetic energy
flux of material with a mass density of $2 \times 10^{-26}\,{\rm
g\,cm^{-3}}$ flowing through a spherical surface of radius 10 kpc with
a radial velocity of $10^{8}\,{\rm cm\,s^{-1}}$. Thus, the ``cooling
problem'' and our wind speed estimates imply comparable power inputs
from AGN.

The outflows we envisage would probably be somewhat, although not
hugely, collimated. X-ray images of the centre of the Perseus cluster
show some asymmetry in structures at radii approaching 10 kpc. Outside
about 10 kpc, the interaction of AGN plasma with ambient galactic
interstellar and intracluster gas appears to reduce asymmetries
associated with collimated AGN outflows. The filaments extend several
times this far in radii, and are radial suggesting that the wind may
be fairly isotropic at such radii.

One would also expect the outflows to be somewhat
time-dependent. However, the presence of clouds dotted throughout the
ICM may well have the effect of obstructing the outflow of energy
injected by the AGN. Therefore, the time-variability of the AGN-driven
flow far away from its source is likely to be fairly uncorrelated with
the variability at the source. Essentially this means that, although
the initial energy injection rate may be time-variable, at larger
radii the variations may have been smoothed-out making the flow
relatively steady. Finally, intermittancy on long timescales much
greater than $3 \times 10^{7}\,{\rm yr}$ will be on timescales greater
than the ages of the structures we are attempting to explain, and
consequently unimportant in this work. Based on these considerations,
we will assume the outflows to be transonic, isotropic and steady.

\section{Numerical setup}
\subsection{Assumptions and methods} 
We use hydrodynamical simulations to investigate the interaction
between a tenuous flow and a mass source. The source of mass is
assumed to be a cold cloud from which material is stripped due to the
interaction with an incident wind. The basic approach is similar to
that adopted by \cite{falle02}, \cite{pittard}, and \cite{dyson}. The
main difference is that here we also include the effects of gravity to
make the model more appropriate for a typical galaxy cluster
environment.

Despite including the gravitational potential, we will ignore
self-gravity for this set of simulations, for the following
reasons. Self-gravity is certainly important in the cold clump at the
bottom of each filament, and would be important in our model if we
were to try to calculate the mass loss rate of the clump. As we have
already stated, the details of the injection process are not important
for the question we are addressing. Self-gravity may well be important
in the filaments, but the absence of a population of bright, young
stars associated with the filaments (e.g. Hatch et al. 2006) also
provides some evidence that it may not. As an illustration, we give
the ratio of the magnitude of the gravitational field to the magnitude
of the thermal pressure force per unit volume in a cylindrical
filament of mass density $\rho$ and radius $R$ is approximately,
\begin{eqnarray}
\frac{g\rho R}{P} = \frac{2\pi G \rho \mu m_{\rm p} R^{2}}{k_{\rm b}T}
\approx 0.3 \bigg(\frac{\rho}{10^{-22}\,{\rm
g\,cm^{-3}}}\bigg)\nonumber \\\bigg(\frac{R}{30\,{\rm
kpc}}\bigg)^{2}\bigg(\frac{T}{10^{4}\,{\rm K}}\bigg)^{-1},
\end{eqnarray}
where $\mu m_{\rm p}$ is the mean mass per particle in the gas.  Thus,
filament parameters are in the range that may allow the ratio to
approach unity. However, it is clear that the values are sufficiently
uncertain that an initial focus on cases in which self-gravity is
neglected is reasonable. Furthermore, if self-gravity does lead to
instability, shear may limit the development of clumps. We will
neglect self-gravity for the present investigation.

The simulations are split into two groups. In the first we assume
constant gravity, and work in the frame of the cloud. This is to gain
an understanding of the effect of the gravitational potential and the
density contrast between the cloud and the wind. In the second group,
we employ a realistic cluster atmosphere and work in the rest frame of
the cluster, and advect the cloud. This allows us to implement what we
have learnt from the prior simulations in the more realistic case. The
latter setup is discussed more fully in section 6. In either case, the
velocity of the incident wind increases, as measured in the clump
frame, with time. Gravity also affects the motion of material in the
tail.

For all models we assume a rate of mass injection per unit time and
volume that may vary temporally, but that is spatially uniform within
a given radius of the cloud's centre. This mass injection prescription
is simple to implement, but has the disadvantage that the flow from
the injection region is isotropic. However, the effect of asymmetric
mass loss is small at large distances from the cloud
\citep[][]{pittard}. Therefore, we emphasize that the actual details
of the mass injection process are relatively unimportant. The boundary
of the mass injection region is not meant to coincide with the
boundary of a cloud, and the cloud could be much smaller than the
injection region.

Three-dimensional calculations are required if the sources are
spherical. To reduce the computational cost we restrict ourselves to
two-dimensional simulations in which the sources are cylindrical. While
we expect some differences between calculations performed in two and
three dimensions, at this stage we can still gain important insight
from less computationally demanding two-dimensional simulations.

In this study we treat the hot gas as adiabatic, since in the Perseus
cluster the cooling time of X-ray emitting gas at 30 kpc is $\sim
10^{9}\,{\rm yr}$, much longer than the flow time of the outflow along
the tail. The cooling rate of the gas in the tail itself is very high,
and must obviously be offset by a heating rate. We do not know the
precise temperature of this gas, but the optically emitting gas in the
tail is at about $10^{4}$K. An isothermal treatment of this material
is reasonable. Thus, we investigate the simple case in which the
incident wind behaves adiabatically, while the injected gas remains
isothermal.

To ensure isothermal behaviour, we use an advected scalar, $\alpha$,
which is unity in the injected gas and zero in the ambient gas. The
source term in the energy equation is then
\begin{equation}
\kappa\alpha\rho(T_{0}-T),
\end{equation}
where $\rho$ and $T$ are the local mass density and temperature, and
where $\kappa$ is large enough that the temperature always remains
close to the equilibrium temperature, $T_0$, in the injected
gas. Inside the injection region we added an extra energy source so
that the gas is injected with temperature $T_{0}$ \citep[see][for
further details]{falle02}.

The simulations are performed using $mg\_gt$, a second-order accurate
code with adaptive mesh refinement (AMR). $mg\_gt$ uses a hierarchy of
grids $G^{0}-G^{\rm N}$ such that the mesh spacing on grid $G^{\rm n}$
is $\Delta x_{0}/2n$. Grids $G^0$ and $G^1$ cover the whole domain,
but the finer grids only exist where they are needed. The solution at
each position is calculated on all grids that exist there, and the
difference between these solutions is used to control refinement. In
order to ensure Courant number matching at the boundaries between
coarse and fine grids, the time-step on grid $G^{\rm n}$ is $\Delta
t_{0}/2n$ where $\Delta t_0$ is the time-step on $G^0$.

In this model, we are studying the behaviour of cold clouds located
$\sim 10$ kpc away from the centre of the cluster. At such distances,
the gravitational acceleration in galaxy clusters is $ \sim
10^{-7}{\rm cm\,s^{-2}}$. We investigate different density contrasts
between the cloud material and the ambient gas, and compare the
morphology of the results with observations.

\subsection{Variables}

The problem we are investigating is scale-free. For numerical reasons,
we have chosen scales so that the values of parameters are near
unity. For easier comparison with the astrophysical systems of
interest we will rescale the results to physical units. In particular,
we have scaled the length by a factor $L$, the gravitational
acceleration by $G$, and the density by $R$. In this case, the
hydrodynamic quantities scale as,
\begin{equation}
x = Lx',
\end{equation}
\begin{equation}
g = Gg',
\end{equation}
\begin{equation}
\rho = R\rho',
\end{equation}
\begin{equation}
t = \tau t'
\end{equation}
\begin{equation}
v = \frac{L}{\tau}v',
\end{equation}
\begin{equation}
p = R\bigg(\frac{L}{\tau}\bigg)^{2}p',
\end{equation}
where the primed quantities represent the dimensionless simulation values,
and the unprimed quantities represent the real values. $t$ is
time, $v$ is velocity, and $p$ is pressure. In cases in which $g$ is
nonzero
\begin {equation}
\tau = \sqrt{\frac{L}{G}}
\end {equation}
In our simulations the values selected for parameters are roughly
consistent with estimates of the properties of clouds in the Perseus
cluster: the length $L$ = 50 pc, $G = 10^{-7}{\rm cm\,s^{-2}}$ (except
for one case) and $R = 10^{-23}{\rm g\,cm^{-3}}$. For cases in which
$g$ = 0, we leave $\tau = 1.24$ Myrs, which is the same value as when
$g$ is nonzero. In section 5, we have employed more accurate values of
the scaling factors.

\subsection{Initial conditions}

To ensure that the ambient wind does not penetrate the injection
region, we take the ram pressure of the injected material
at the boundary of the injection region to be larger than that of the
wind, i.e. we require,
\begin{equation}
{v_{\rm i}}^{2}\rho_{\rm i} = \frac{v_{\rm i}r_{\rm i}Q}{3} \geq
\rho_{\rm w}v_{\rm w}^{2},
\end{equation}
where $r_{\rm i}$ is the radius of the injection region, $v_{\rm i}$
is the flow speed of the injected material at $r = r_{\rm i}$ and
$\rho_{\rm i}$ its density at this point. $Q$ is the rate per unit
volume per unit time at which mass is injected within $r=r_{\rm i}$.

If we use the pressure behind a stationary normal shock in the wind
rather than the wind's ram pressure, the above criterion becomes
\begin{equation}
  Q \ge \frac{4}{\gamma + 1}\frac{\rho_{\rm w}v_{\rm
  w,0}^{2}}{v_{\rm i}r_{\rm i}}.
\end{equation}

In the frame of a falling clump, the wind's ram pressure increases
with time. Thus, the minimum mass injection rate must increase
accordingly to ensure that the wind does not penetrate the injection
region. If the value of $Q$ at $t=0$ is $Q_{0}$ and the gravity is
constant, the ratio of $Q$ to the wind ram pressure will remain
constant if
\begin{equation}
Q = Q_{0}\bigg(1+ \frac{gt}{v_{\rm w,0}} \bigg)^{2},
\end{equation}
where $v_{\rm w,0}$ is the relative velocity between the clump and the
ambient flow at $t=0$. While the method that we are using to treat
mass injection requires that we take $Q$ to increase with time, in
reality the mass injection rate will increase with ram pressure
\citep[e.g.][]{hart}.

The computational domain is $-200 \le x \le 200$, $-100 \le y \le
400$, with the mass injection region centred at the origin. We use 6
grids, and the size of the finest cells is 0.125.

By scaling the physical variables we can choose units such that
$r_{\rm i}=1$, $v_{\rm i}=1$, and $\rho_{\rm w}=1$ and initially set
$v_{\rm w} = 10$ and $T_{\rm 0}=1$. Furthermore, we assume that the
injected material is in pressure equilibrium with the ambient
wind. Thus, if we assume that the injected material is a factor of
$10^3$ cooler than the ambient wind, then it is also $10^3$ times
denser than the wind. We will use $\eta$ to represent this contrast
parameter.

We investigate the morphology of tails for three different contrast
parameters: $\eta = 10^2$, $10^3$, and $10^4$. The corresponding
initial values of the adiabatic Mach number, $M$, are: 0.77, 0.24, and
0.077, respectively. These parameters define case 1), 2), and 3),
respectively. Clearly, as the ambient wind accelerates with respect to
the cloud, the Mach number will increase.

For comparison, we also study several other cases in which the density
contrast is $10^3$. Case 2i) is identical to case 2) except that $g'=
10$ and $G=10^{-8}$. In cases 2ii) and 2iii) the gravity is zero, but
there is a constant wind velocity of $10$ and $30$ (in code units)
respectively, corresponding to Mach numbers of 0.24 and 0.72. These
parameters are summarised in table 1. It does not make sense to
specify the mass of the clouds, since they are likely to be composed
partly of molecular material, and therefore considerably more massive
than an estimate of $4\pi r_{\rm c}^{3}\rho_{\rm c}/3$ would suggest.

\begin{table*}
\centering
\begin{minipage}{140mm}
\caption{Simulation parameters: ambient density, cloud density, sound
speed of injected material, initial Mach number of the wind velocity
with respect to cold material, gravitational acceleration. The implied
mass outflow rates close to the origin are $4\pi r^{2}\delta \rho_{\rm
amb}v_{\rm w} \sim 10\,{\rm M_\odot\,yr^{-1}}$, where $\delta \sim
0.1$ is the fraction of the sky covered by the outflow.}
\begin{tabular}{lccccc}
\hline Case & $\rho_{\rm amb}\,(g\,cm^{-3})$ & $\rho_{\rm c}\,(g\,cm^{-3})$ & $c_{\rm s,c}\,({\rm cm\,s^{-1}})$ & $M_{\rm initial}$ & $g\,({\rm cm\,s^{-2}})$\\ 
\hline 
1 & $10^{-26}$ & $ 10^{-24}$ & $5.2 \times 10^{5}$ & 0.77 & $10^{-7}$\\
2 & $ 10^{-26}$ & $ 10^{-23}$ & $1.6 \times 10^{5}$ & 0.24 & $10^{-7}$\\ 
3 & $ 10^{-26}$ & $ 10^{-22}$ & $5.2 \times 10^{4}$ & 0.077 & $10^{-7}$\\ 
2i & $ 10^{-26}$ & $ 10^{-23}$ & $1.6 \times 10^{5}$ & 0.24 & $10^{-7}$\\ 
2ii & $ 10^{-26}$ & $ 10^{-23}$ & $1.6 \times 10^{5}$ & 0.24 & $10^{-7}$\\ 
2iii & $10^{-26}$ & $ 10^{-23}$ & $1.6 \times 10^{5}$ & 0.72 & $10^{-7}$\\ 
\hline
\end{tabular}
\end{minipage}
\label{table1}
\end{table*}

The code wind velocity $v_{\rm w}' = 10$, corresponds to a velocity of
$v_{\rm w} = 3.9 \times 10^{7}\,{\rm cm\,s^{-1}}$, which, is
comparable with the transonic values described in the previous
section.

\subsection{Sub-grid turbulence model}
Simulations with a transonic wind pose difficulties that are absent in
those for hypersonic winds. Firstly, in order for the boundaries to
have no effect on a simulation with a transonic wind, the
computational domain has to be very large.  Secondly, the velocity
shear between the injected material and the wind is so extreme in the
transonic case that the wind flow separates and produces a turbulent
wake downstream of the interaction region. Calculations based on the
Euler equations cannot adequately describe such turbulence, and a
turbulence model must be employed.

We adopt the $k-\epsilon$ model used by \cite{falle02}.  The subgrid
turbulence model allows the simulation of a high Reynolds number flow
through the inclusion of equations for the turbulent energy density
and dissipation rate, from which viscous and diffusive terms are
calculated. The model solution gives an approximation to the mean
flow. The diffusive terms in the fluid equations model the turbulent
mixing of the injected gas with the original flow due to shear
instabilities.  The model has been calibrated by comparing the
computed growth of shear layers with experiments (Dash \& Wolf
1983). It is based on the assumptions that the real Reynolds number is
very large, which is the case in astrophysical flows, and that the
turbulence is fully developed. The use of the subgrid model gives more
realistic results than those obtained from a hydrodynamical code where
the sizes of the shear instabilities are determined by the numerical
resolution. Further details can be found in \cite{falle94}.

\section{Results}

\subsection{Morphology}

\begin{figure}
\centering
\includegraphics[width=5cm]{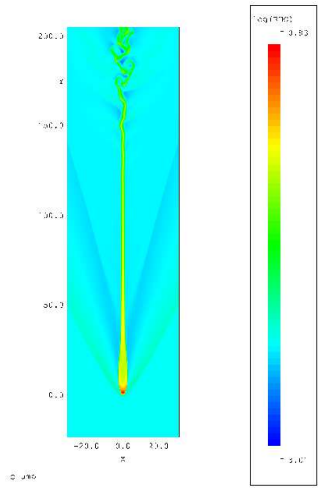}
\caption{Density morphology for case 1) at $t'\approx 28$ ($t \approx
  35$ Myrs) at which time the external flow at the position of the
  clump has accelerated to $M$ = 3. The initial Mach number was
  0.77. Each spatial unit corresponds to 50 pc. Thus, the plotted
  domain, in the x-direction, extends between $\pm$ 1.5 kpc. The
  density contrast between the cold clump and the ambient gas is
  $10^2$. Note the bow shock around the clump, and also the tail shock
  behind this.}
\label{fig:dens1}
\end{figure}

\begin{figure}
\centering
\includegraphics[width=5cm]{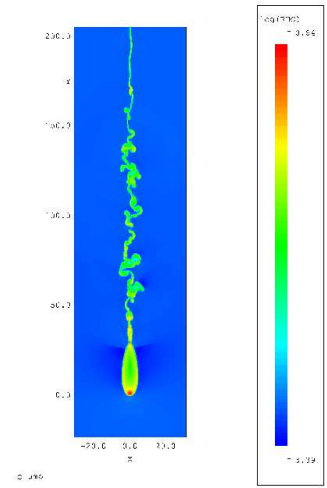}
\caption{Density morphology for case 2) at $t' \approx 17$ ($t \approx
21$ Myrs) when $M$ = 0.66. The initial Mach number was 0.24. The
density contrast between the cold clump and the ambient gas is
$10^3$. Note the irregular structure of the tail along the entire
length of the filament, compared to the linear tail in the previous
figure.}
\label{fig:dens2}
\end{figure}

\begin{figure}
\centering
\includegraphics[width=5cm]{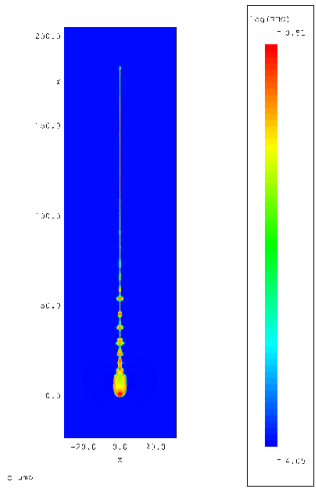}
\caption{Density morphology for case 3) at $t' \approx 18$ ($t \approx
22$ Myrs) when $M$ = 0.22. The initial Mach number was 0.077. The
density contrast between the cold clump and the ambient gas is
$10^4$. Note the approximately linear tail, with at least 6 descrete
blobs of material close to the clump.}
\label{fig:dens3}
\end{figure}

\begin{figure}
\centering
\includegraphics[width=5cm]{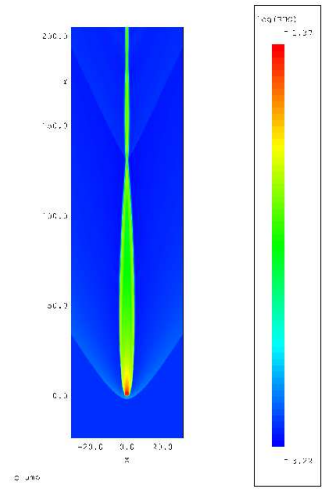}
\caption{Density morphology for case 2i) at $t' \approx 20$ ($t
\approx 25$ Myrs) when $M$ = 5.2. The initial Mach number is 0.24. The
density contrast between the cold clump and the ambient gas is
$10^3$. The structure of the tail in this example is regular, with no
obvious blobs at various locations along the tail.}
\label{fig:dens4}
\end{figure}

In general, the morphologies of the filament density generated by
accelerating flows are very similar to those presented in
\cite{pittard}. A bow wave develops upstream of a mass source, and the
wave's amplitude falls off as $1/r$. If the flow becomes supersonic,
the bow wave becomes a bow shock and a very weak tail shock may occur
in the wind downstream of the mass source, when the external flow is
subsonic. The injected material remains in approximate pressure
equilibrium with the wind, and eventually becomes confined to a tail
with a width comparable to that of the injection region (cf. figures
\ref{fig:dens1} to \ref{fig:dens4}). The opening angle of the injected
material increases with the external Mach number.

It should be noted that we use a sub-grid turbulence model which, in
effect, models a turbulent boundary layer at the surface of the
filament. So even when the tail appears smooth there is significant
turbulence. For increasingly large density contrasts, the turbulent
viscosity, which is an integral part of the subgrid turbulence model,
fails to stabilize the subsonic shear layer. In this case, there is
turbulence on the grid as well as sub-grid turbulence. Large-scale
instabilities in thin filaments demand high density contrasts and an
ambient flow that remains subsonic with respect to the clump.

The opening angle of a turbulent boundary layer is given by Canto \&
Raga (1991) as,
\begin{equation}
\frac{{\rm d}h}{{\rm d}x} = \frac{2}{\sigma}
\end{equation}
where $\sigma$ is the so-called spreading parameter. Experiments show
that at low Mach number, turbulent mixing layers have a constant
opening angle, and $\sigma \sim 11$, so that ${\rm d}h/{\rm d}x \sim
10.4$. At high Mach numbers, $\sigma \sim 50$, so ${\rm d}h/{\rm d}x
\sim 2.3$. The tail becomes fully turbulent at a downstream distance
of,
\begin{equation}
x = \frac{r_{\rm c}M_{\rm w}}{2\epsilon}\frac{c_0}{c_{\rm w}}.
\end{equation}
where $\epsilon$ is the entrainment efficiency, $M_{\rm w}$ is the
Mach number of the wind with respect to the clump, $c_{s,a}$ is the
sound speed of the environment, $c_{\rm s,c}$ is the sound speed of
the tail material, $r_{\rm c}$ is the initial radius of the tail. 

To calculate this length, Canto \& Raga use $x = 16.9 M_{\rm w}r_{\rm
c}$, which can be rescaled as,
\begin{equation}
x \approx 0.85\,{\rm kpc}\bigg(\frac{M_{\rm
w}}{1}\bigg)\bigg(\frac{r_{\rm c}}{50\,{\rm pc}}\bigg),
\end{equation}
in other words, for sensible values of the wind flowing past a clump,
we should expect the tail to be turbulent on kpc scales.

\begin{figure}
\centering
\includegraphics[width=5cm]{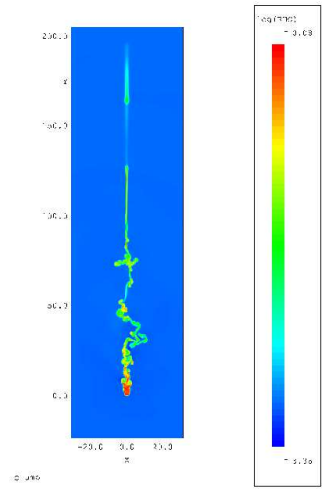}
\caption{Density morphology for case 2ii) at $t' \approx 20$ ($t
\approx 25$ Myrs). The density contrast between the cold clump and the
ambient gas is $10^3$, and gravity is omitted from the simulation. $M$
= 0.24.}
\label{fig:dens5}

\end{figure}

\begin{figure}
\centering
\includegraphics[width=5cm]{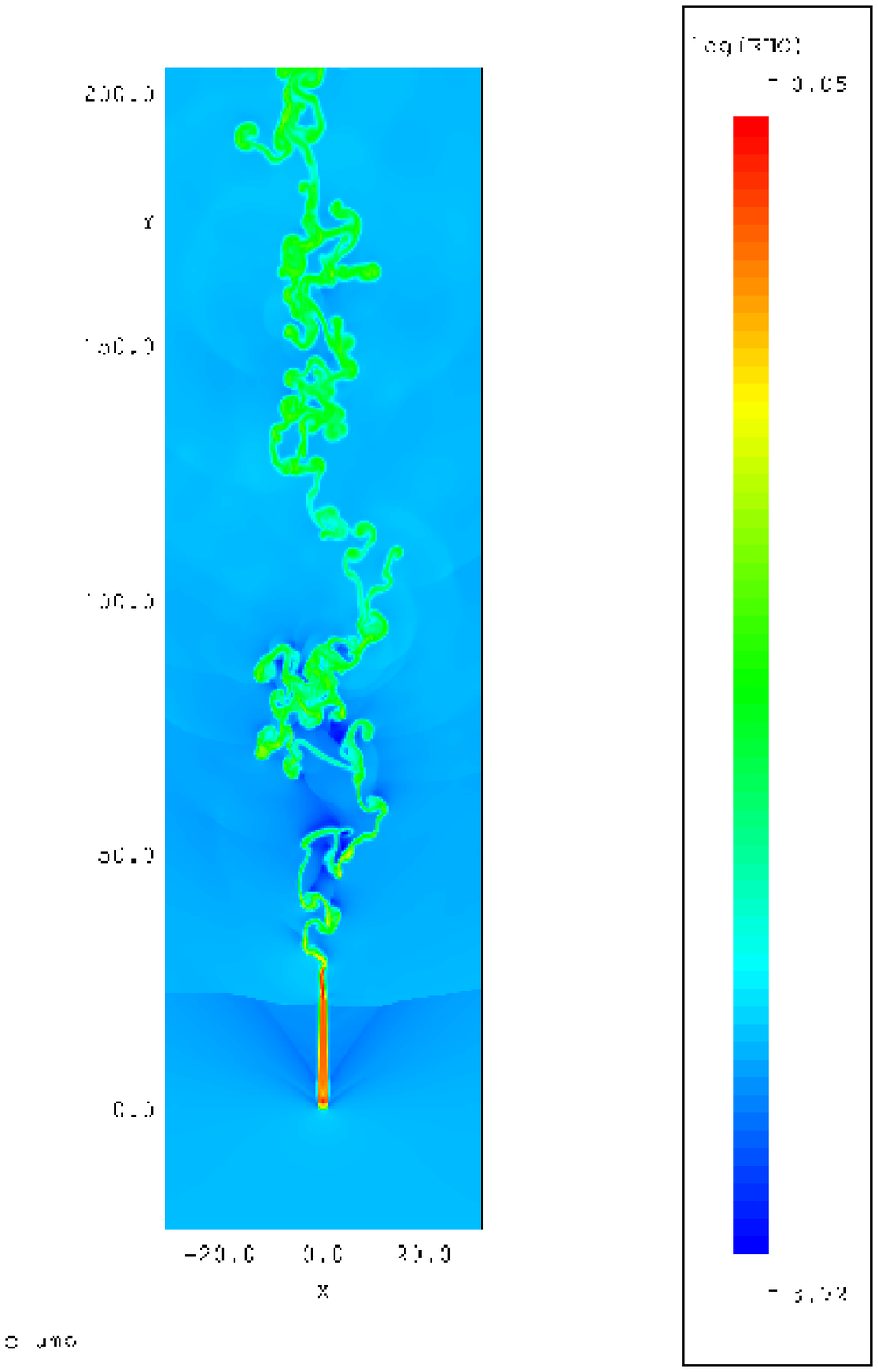}
\caption{Density morphology for case 2iii) at a code time of $t'
\approx 20$ ($ t\approx 25$ Myrs). The density contrast between the
cold clump and the ambient gas is $10^3$. Gravity is omitted from this
simulation. $M$ = 0.72.}
\label{fig:dens6}
\end{figure}

These figures show a clear dependence, at least at early times, of the
tail morphology on the density contrast between the cloud and the
ambient gas. For example, for $\eta = 10^{4}$ (figure 3), the tail is
nearly linear, but is punctuated by fairly evenly spaced blobs along
the length of the tail. This is because subsonic shear layers tend to
be unstable unless the viscosity is high. Some turbulent jets display
similar behaviour. The blobs are separated by increasingly large
distances further along the tail, indicating acceleration. For $\eta
=10^2$, (figure 1) the tail only becomes irregular at large distances
away from the cloud, while for $\eta=10^3$ (figure 2) most of the tail
is irregular. For comparison we also show the morphologies for a case
when the gravity is weaker (figure 4) and nongravitational cases with
different ambient flow velocities (figures \ref{fig:dens5} and
\ref{fig:dens6}).


For ambient flows that remain subsonic in the frame of the clumps, the
tails are irregular in appearance. In case 2i) the gravity is strong
enough that the initially subsonic flow becomes supersonic, and
consequently the tail becomes smooth and regular. Morphological
considerations by themselves would suggest that the filaments in the
Perseus cluster satisfy these conditions. However, an examination of
the kinematic structures of the model and observed filaments is also
necessary.

\subsection{Velocity profiles}

\begin{figure*}
\centering
\includegraphics[width=14cm]{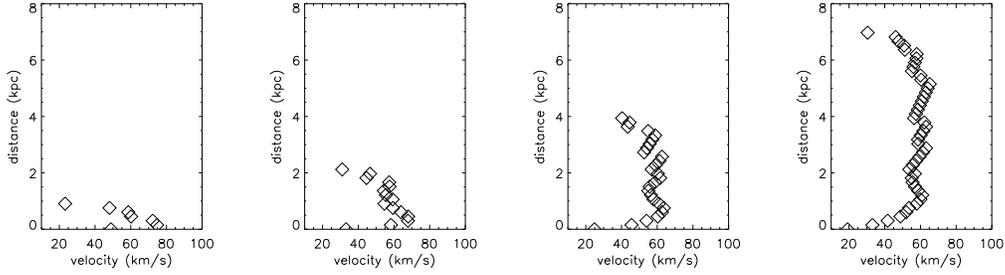}
\caption{Velocity profile for case 1) for (from left to right) $t'=$
  6.95, 14.5, 20.9, and 28.2 ($t=$ 8.3, 18, 26, and 35 Myrs),
  respectively. The Mach numbers of the ambient flow with respect to
  the cold clump are: 1.3, 1.9, 2.4, and 3, respectively. We have
  removed the central 10 $\rm km\,s^{-1}$ to avoid any points where
  the tail does not exist.}\label{fig:vel1}
\end{figure*}

\begin{figure*}
\centering
\includegraphics[width=14cm]{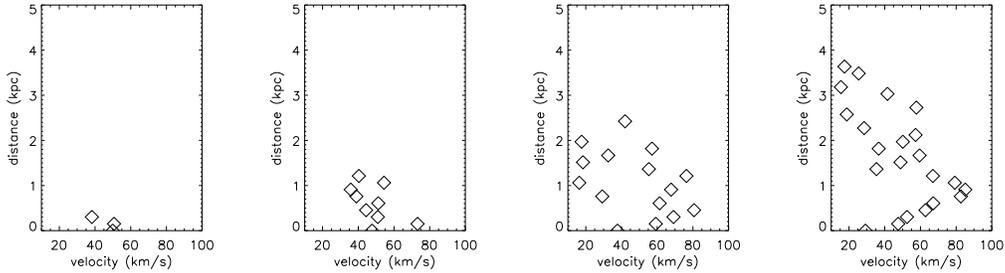}
\caption{Velocity profile for case 2) for $t'=$ 3.4, 8.3, 12.1, and
  16.9 ($t=$ 4.3, 10, 15, and 21 Myrs) respectively. The Mach numbers
  of the ambient flow with respect to the cold clump are 0.33, 0.45,
  0.54 and 0.66, respectively. We have removed the central 10 $\rm
  km\,s^{-1}$ to avoid any points where the tail does not
  exist.}\label{fig:vel2}
\end{figure*}

\begin{figure*}
\centering
\includegraphics[width=14cm]{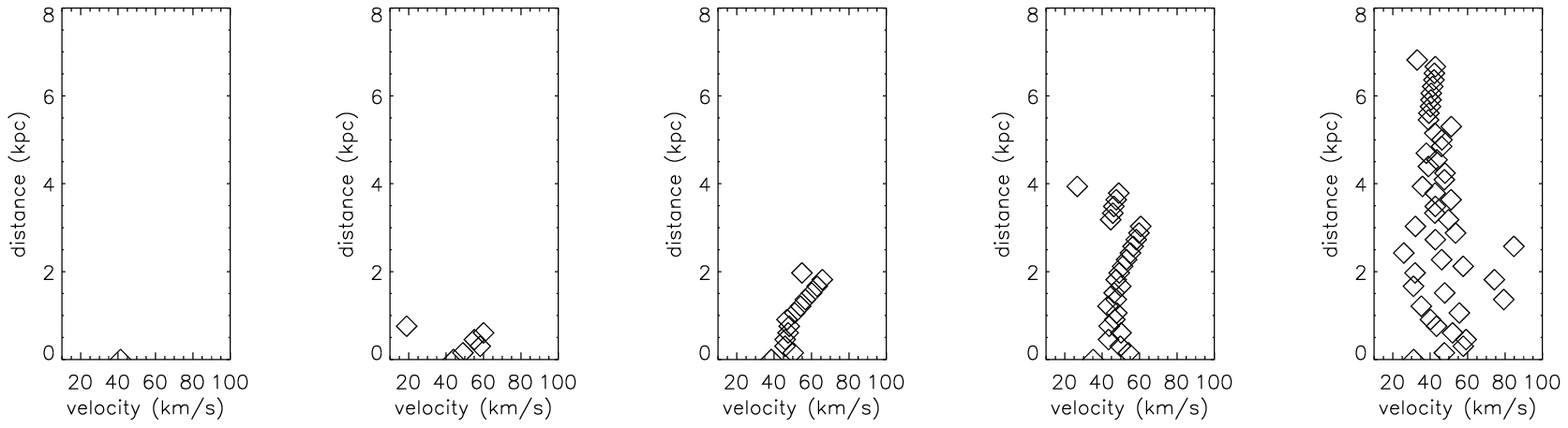}
\caption{Velocity profile for case 3) for $t'=$ 1.93, 5.7, 12.7 and
  18.1 ($t=$2.4, 7.0, 12, and 22 Myrs) respectively. The Mach numbers
  of the ambient flow with respect to the cold clump are 0.09, 0.12,
  0.15, 0.18 and 0.22, respectively. We have removed the central 10
  $\rm km\,s^{-1}$ to avoid any points where the tail does not
  exist.}\label{fig:vel3}
\end{figure*}

\begin{figure*}
\centering
\includegraphics[width=14cm]{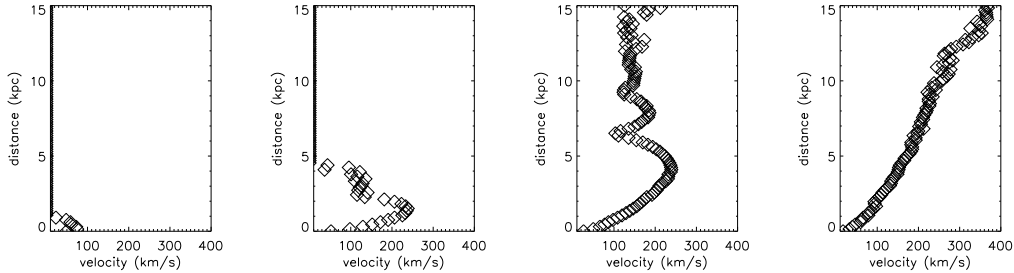}
\caption{Velocity profile for case 2i) at $t'=$ 2, 5, 10, and 20 ($t=$
  2.5, 6.2, 12, and 25 Myrs), respectively. The Mach numbers of the
  ambient flow with respect to the cold clump are 0.74, 1.5, 2.7 and
  5.2, respectively. We have removed the central 10 $\rm km\,s^{-1}$
  to avoid any points where the tail does not exist.}\label{fig:vel4}
\end{figure*}

Figures \ref{fig:vel1} to \ref{fig:vel4} show the mean velocity of the
tail material at 100 increments along the length of the tail for each
of cases 1), 2), 3), and 2i).

Except for case 2i) and to some extent case 1), each velocity profile
shows some scatter around a mean velocity that is roughly constant
along the length of most of the filament.  We do not show velocity
profiles for cases 2ii) and 2iii) because they show behaviours similar
to those for cases 2) and 3).  However, the velocity profiles,
including those for the model filaments that bear the most
morphological resemblence to the observed filaments, are not
compatible with the observed velocity structures of the NGC 1275
filaments. Nevertheless, the third panel of figure \ref{fig:vel4}
shows clear zig-zag features, which are similar to those in the
velocity profiles given in figures 5, 7 and 14 of \cite{hatch}. Note
that the zig-zag features in figures \ref{fig:vel1} and \ref{fig:vel4}
also correspond to the regions of enhanced density along the tail. The
similarity between figures \ref{fig:vel1} and \ref{fig:vel4} suggests
that the zig-zag features occur when the Mach number of the ambient
flow, with respect to the clump, is roughly 2-3, after $\sim$ few 10s
Myrs.

The velocities near the heads of the model filaments are low, whereas
those of the observed filaments are not. The discrepency could be due
to the material near the heads of the observed filaments being warm,
but neutral, and hence invisible in the observed optical lines. More
likely, it is a result of our prescription for mass injection which
will produce different features near the injection region, but the
large-scale downstream flow should be unaffected.

The fact that the profile shown in figure \ref{fig:vel4} resembles the
observed profile better than that shown in figure \ref{fig:vel1}, is
probably because the density contrast between the cold clump and the
ambient flow is more realistic in case 2i). It is likely that if we
had continued simulating cases 2) and 3) to high enough Mach numbers
that they may also have begun to exhibit the zig-zag structure. Based
on these results it appears that the velocity fluctuations (and
scale-size of instabilities) are larger for greater density contrasts
between the clump and the wind.

\section{Tails in a realistic cluster environment: the Perseus cluster}

Having investigated the properties of accelerating flows past cold
clouds, we now study the more specific case of a galaxy cluster
atmosphere. Firstly, we assume a generic gravitational potential which
leads to a $\beta$-profile gas density profile in hydrostatic
equilibrium \citep[e.g.][]{vern05},
\begin{equation}
\rho = \frac{\rho_{0}}{[1 + (r/r_{0})^{2}]^{\beta}},
\end{equation}
where we have assumed spherical symmetry, $r$ is the radial
coordinate, and $r_{0}$ is the scale-height of the density
distribution. The gravitational acceleration is therefore given by
\begin{equation}
g = -2 \beta \frac{k_{\rm b}T}{\mu m_{\rm p}}\bigg(\frac{r}{r_{0}^{2}
+ r^{2}}\bigg),
\end{equation}
where $T$ is the gas temperature which is assumed to be constant with
radius, $\mu = 0.6$ is the mean mass per particle.

For each such model, we will perform the calculations in the frame of
the central galaxy and prescribe how the cloud injection region
moves. Due to the structure of the gravitational potential employed
and the initial location assumed for the cloud, the solution of the
cloud motion is very close to that for free-fall. Assuming free-fall,
we find that the velocity, after starting from rest at $x_{0}$, is,
\begin{equation} \label{eq:vofr}
v(r) = -\bigg(2 \beta \frac{k_{\rm b}T}{\mu m_{\rm
p}}\bigg)^{1/2}\bigg[\frac{r_{0}^{2}+x_{0}^{2}}{r_{0}^{2}+r^{2}}\bigg]^{1/2}.
\end{equation}

Equation (\ref{eq:vofr}) can be numerically integrated, using the
values for specific galaxy clusters, to obtain the position of the
cloud at each point in time, $r(t)$. For example, for the Perseus
cluster we use, $T = 10^{8}\,$K, $\rho_{0} = 7 \times 10^{-26}\,{\rm
g\,cm^{-3}}$, $\beta = 0.81$, and $r_{0} = 28.5\,$kpc
\citep[e.g.][]{pope06}. With these values, the motion of a cloud
falling freely from $30\,$kpc can be well approximated by,
\begin{equation}
r(t) = 30\bigg[1 - 0.9\bigg(\frac{t}{75\,{\rm
Myr}}\bigg)^{2}\bigg]\,{\rm kpc},
\end{equation}
which is roughly what one would expect for constant gravity. This
correspondence occurs because the gravitational acceleration scales as
$1/r$ at large radii, and $r$ at small radii. The cloud is initially
located around the turning point where the function is relatively
flat. This means that constant gravity is a suitable approximation and
the previous set of simulations, in which constant gravity was
assumed, retain their validity.

For completeness, the Mach number of the flow around the clump, with
respect to the ambient flow, is
\begin{equation}
M =
\frac{2\beta}{\gamma^{1/2}}\bigg(\frac{r_{0}^{2}+x_{0}^{2}}{r_{0}^{2}+r^{2}}\bigg).
\end{equation}
Thus, for $\beta = 3/2$, $r_0 = 28.5$, $x_0 = 30\,$kpc, we find that
$M > 1$ for $r \lesssim 0.9 r_0 \approx 26\,$kpc. Consequently, for
higher wind velocities, the flow becomes supersonic closer to the
initial location, $x_0$. The ram pressure due to relative motion
between the cloud and the static atmosphere is,
\begin{equation}
\rho_{\rm w}v_{\rm w}^{2} = \rho[r(t)]\bigg[\frac{{\rm d}r(t)}{{\rm
d}t}\bigg]^{2}.
\end{equation}

We choose an initial density contrast between the cloud and its
surrounding of $10^{4}$, and set the scaling parameters to $R = 7
\times 10^{-22}\,{\rm g\,cm^{-3}}$ and $L = 3.085 \times 10^{21}\,{\rm
cm}$. $P = \rho k_{\rm b}T/\mu m_{\rm p} \approx 3.6 \times 10^{-10}\,
{\rm erg\,cm^{-3}}$ in the centre of the Perseus cluster, giving $G =
P/(P'RL)= 1.66 \times 10^{-10}\, {\rm cm\,s^{-1}}$. The temporal
scaling constant, is $\tau = 4.3 \times 10^{15}\,{\rm s}$, while the
velocities are scaled by $L/\tau = 7.2 \times 10^{5}\,{\rm
cm\,s^{-1}}$. The time taken for the cloud to fall from 30 kpc to 3
kpc is 75 Myr, or 0.55 in terms of code time units.  For a density
contrast of $10^{3}$, we use $R = 7 \times 10^{-23}\,{\rm g\,cm^{-3}}$
and $L = 3.085 \times 10^{21}\,{\rm cm}$, giving $G = 1.66 \times
10^{-9}\, {\rm cm\,s^{-1}}$. This means that the temporal scaling
constant becomes $\tau = 1.4 \times 10^{15}\,{\rm s}$, and the
velocities are scaled by $L/\tau = 2.2 \times 10^{6}\,{\rm
cm\,s^{-1}}$. The time taken for the cloud to fall from 30 kpc to 3
kpc now, or 1.7 in terms of code time units.

Finally, in the preliminary simulations we found that the tail widths
were comparable to the diameters of the clouds. Therefore, to generate
filaments of the observed widths which are constrained to be $<$1 kpc
we will simulate the effect of material stripped from relatively large
clouds, with an injection region of radius 100 pc. As already stated
this may be significantly larger than the physical size of the cloud.

\subsection{Results}

Since the lateral velocity dispersion (zig-zag effect) appears to
increase for larger density contrasts, we will concentrate on the case
$\eta=10^{4}$. However, the case where $\eta=10^{3}$ is also of
interest. The general properties are as follows: long tail-like
structures only form if the wind velocity is sufficiently large. This
wind is assumed to be the same material, at the same temperature as
the static atmosphere. For example, for low values of $v_{\rm w}'$ (5
and 10, corresponding to $3.6\times 10^{6}$ and $7.2\times
10^{6}\,{\rm cm\,s^{-1}}$ for $\eta =10^{4}$), the tail never becomes
long because the ram pressure due to the impinging wind cannot
overcome the gravitational force acting on the tail material. As a
result any tail that forms is infalling everywhere, which contradicts
the observational data, at least for NGC 1275. It therefore appears
that the length and morphology of the tails are extremely dependent on
the wind velocity.

\begin{figure}
\centering
\includegraphics[width=5cm]{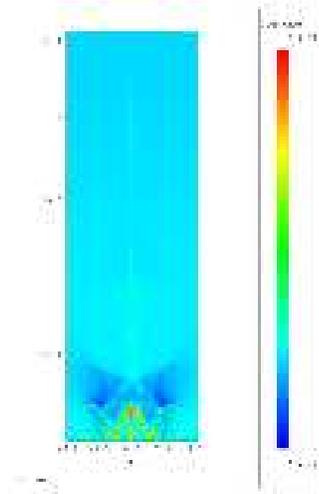}
\caption{The density structure at 58 Myrs, for $\eta = 10^{4}$ and
 wind velocity = $5.8 \times 10^{7}\,{\rm cm\,s^{-1}}$. Panel shows
 data at $\, 3.4 \times 10^{7}\,{\rm yrs}$. The length scale indicates
 the distance from the cluster centre in kpc. The density scale shows
 the density in code units ($\rho'$).}
\label{fig:mor1}
\end{figure}

\begin{figure}
\centering
\includegraphics[width=5cm]{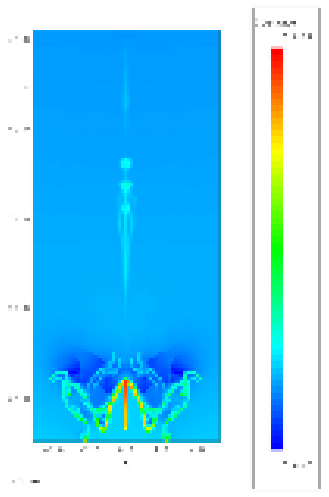}
\caption{The density structure at 38 Myrs, for $\eta = 10^{4}$ and
wind velocity = $4.32 \times 10^{8}\,{\rm cm\,s^{-1}}$. The length
scale indicates the distance from the cluster centre in kpc. The
density scale shows the density in code units ($\rho'$).}
\label{fig:mor2}
\end{figure}

\begin{figure}
\centering
\includegraphics[width=5cm]{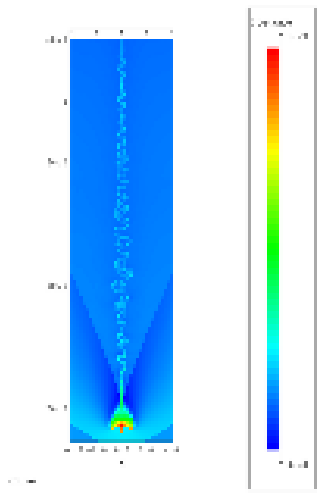}
\caption{The density structure at 20 Myrs, for $\eta = 10^{3}$ and
 wind velocity = $1.76 \times 10^{8}\,{\rm cm\,s^{-1}}$. The length
 scale indicates the distance from the cluster centre in kpc. The
 density scale shows the density in code units ($\rho'$).}
\label{fig:mor3}
\end{figure}

\begin{figure}
\centering
\includegraphics[width=5cm]{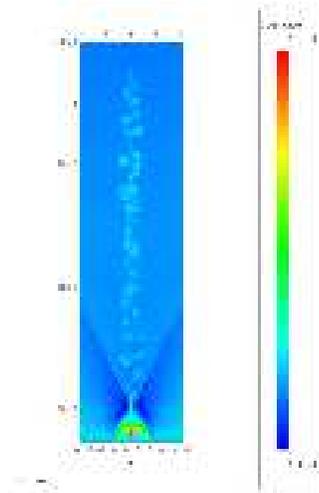}
\caption{The density structure at 24 Myrs, for $\eta = 10^{3}$ and
 wind velocity = $1.32 \times 10^{8}\,{\rm cm\,s^{-1}}$. The length
 scale indicates the distance from the cluster centre in kpc. The
 density scale shows the density in code units ($\rho'$).}
\label{fig:mor4}
\end{figure}

The morphology of the tail also depends strongly on the density
contrast between the cloud and its environment. In each figure, the
cloud is shown as red and is the lowest point along the line defined
by x=0. Material is injected (stripped) and collects behind the
free-falling cloud, and some dense material sinks either side of the
cloud. Behind this cold material, a filamentary tail
forms. Inhomogeneities in the tails are dependent on the wind speed
and density contrast, in agreement with the previous results. As can
be seen from figures \ref{fig:mor1} to \ref{fig:mor4}, it appears that
a density contrast of $\eta=10^{3}$ produces structures most
qualitatively similar to those seen in real filaments. However, they
are low density, being a few $\times$ 0.1 particles per cubic
centimetre. This does not present problems in terms of pressure
balance between the tail and the ambient medium since the tail
material is diluted and heated to temperatures in the range
$10^{5}-10^{7}$K. The simulations produce filaments of
$10^{6}-10^{7}\,{\rm M_\odot}$ while the estimates in the following
section suggest that $10^{8}\,{\rm M_\odot}$ filaments might be more
likely. Even so, the quantity of cold gas, in the simulated tail, is
more than capable of producing the observed optical luminosity
\citep[see][for example]{hatch07}.

The material around the head of the cloud also exhibits different
behaviour in this set of simulations. Material is stripped from the
cloud by the wind and builds up behind. It appears that in many cases
much of this material does not make it into a tail but simply collects
behind the cloud. Eventually this becomes unstable, probably due to
Rayleigh-Taylor instabilities, and produces the structures seen in the
simulations. Similar structures have been noticed in previous
numerical simulations in the work by \cite{murray04}.

In the simulations, initial wind velocities of $\sim 10^{8}\,{\rm
cm\,s^{-1}}$ are required to produce long tails which are outflowing
at their outer extent. This large velocity is expected, since the
velocity of the wind, at the clump, must exceed the infall velocity of
the clump for material to be carried away from it. By the time the
wind has reached the clump it has done considerable work against
gravity. This leads to the velocity dispersion along the clump of a
few $\times 100\,{\rm km\,s^{-1}}$.

\begin{figure*}
\centering
\includegraphics[width=14cm]{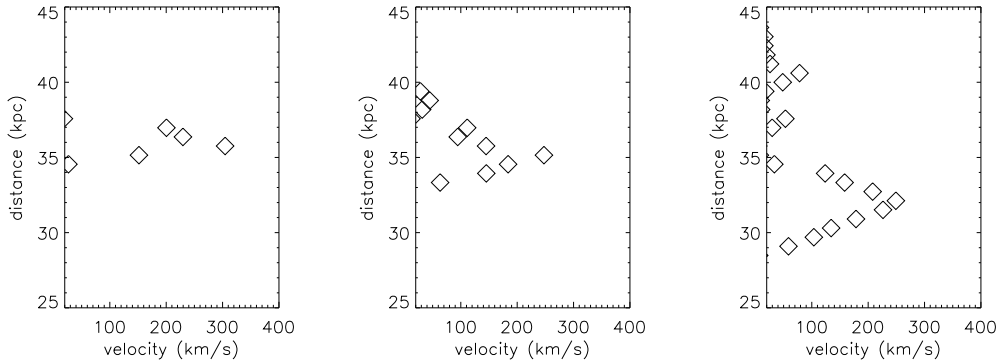}
\caption{Velocity profile for case $\eta = 10^{4}$ and wind velocity =
 $5.8 \times 10^{7}\,{\rm cm\,s^{-1}}$. Panels show data at $14,
 21\,$and$\, 34$ Myrs. The length scale indicates the distance from
 the cluster centre. We have cut off the central 20 $\rm km\,s^{-1}$
 to exclude points where there is no tail.}\label{fig:vel5}
\end{figure*}

\begin{figure*}
\centering
\includegraphics[width=14cm]{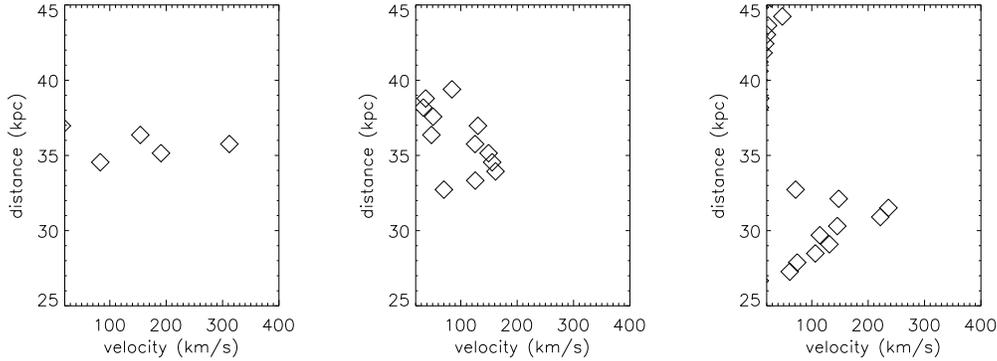}
\caption{Velocity profile for case $\eta = 10^{4}$ and wind velocity =
$4.32 \times 10^{8}\,{\rm cm\,s^{-1}}$. Panels show data at $16, 23\,$
and$\, 38$ Myrs. The length scale indicates the distance from the
cluster centre. We have cut off the central 20 $\rm km\,s^{-1}$ to
exclude points where there is no tail.}\label{fig:vel6}
\end{figure*}

Figure \ref{fig:vel5} shows the velocity profile for $\eta = 10^{4}$
and a wind speed of $5.8 \times 10^{7}\,{\rm cm\,s^{-1}}$ ($v_{\rm
w}'$ = 80). This case probably best resembles the observations. Figure
\ref{fig:vel6} seems to show that a lower wind velocity ($v_{\rm
w}'=60$, $v_{\rm w} =\,4.32\times 10^{7}\,{\rm cm\,s^{-1}}$) does not
produce the observed zig-zag kinematics for this particular density
contrast. These results should be compared with the $\eta = 10^{3}$
case.

\begin{figure*}
\centering
\includegraphics[width=14cm]{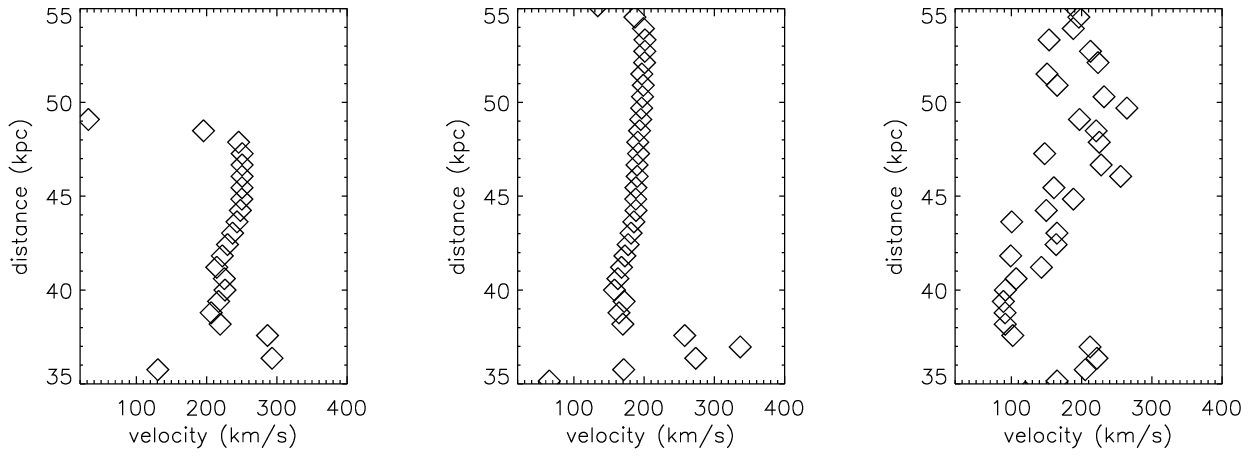}
\caption{Velocity profile for case $\eta = 10^{3}$ and wind velocity =
 $1.76 \times 10^{8}\,{\rm cm\,s^{-1}}$. Panels show data at $8, 12\,$
 and$\, 20$ Myrs. The length scale indicates the distance from the
 cluster centre. We have cut off the central 20 $\rm km\,s^{-1}$ to
 exclude points where there is no tail.}\label{fig:vel7}
\end{figure*}

\begin{figure*}
\centering
\includegraphics[width=14cm]{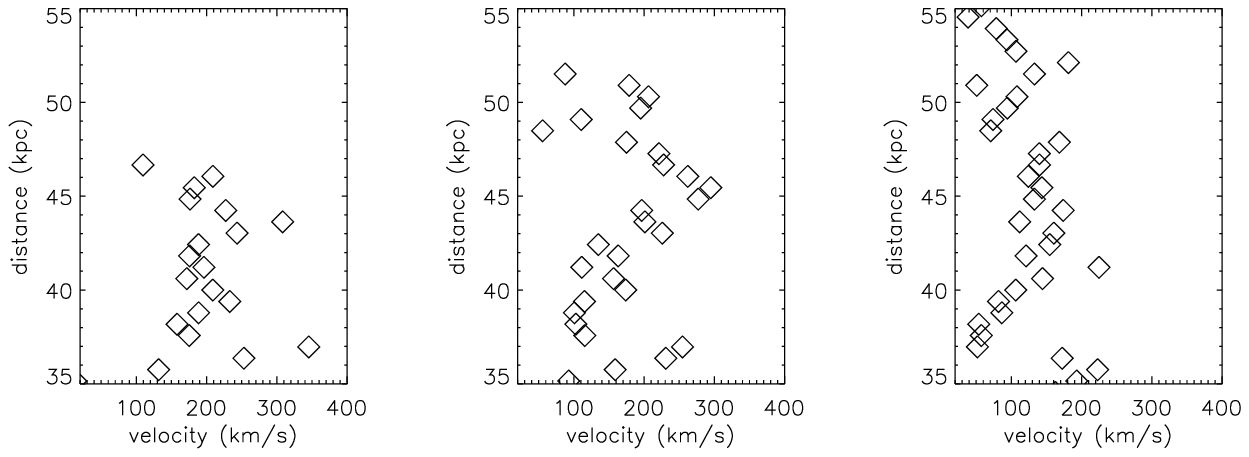}
\caption{Velocity profile for case $\eta = 10^{3}$ and wind velocity =
 $\,1.32 \times 10^{8}\,{\rm cm\,s^{-1}}$. Panels show data at $0.9,
 1.4\,$and$\, 2.4 \times 10^{7}\,{\rm yrs}$. The length scale
 indicates the distance from the cluster centre. We have cut off the
 central 20 $\rm km\,s^{-1}$ to exclude points where there is no
 tail.}\label{fig:vel8}
\end{figure*}

Figures \ref{fig:vel7} and \ref{fig:vel8} show the kinematics for two
wind velocities: $v_{\rm w}' = 80$ and $60$, in figures \ref{fig:vel7}
and \ref{fig:vel8}. These correspond to $1.76 \times 10^{8}\,{\rm
cm\,s^{-1}}$ and $1.32 \times 10^{8}\,{\rm cm\,s^{-1}}$ which are
either side of the sound speed of the ambient medium ($1.52 \times
10^{8}\,{\rm cm\,s^{-1}}$). Figure \ref{fig:vel8} eventually shows
kinematic structure that could be interpreted as a zig-zag, although
it is not as clear as in figure \ref{fig:vel5}. The tails extend to
greater lengths for $\eta = 10^{3}$ than $\eta = 10^{4}$ because the
evaporated material is less dense and can be more easily transported
outwards by the wind.

Again, the amplitude of the zig-zags is smaller for $\eta = 10^{3}$
than $\eta = 10^{4}$. As a result, these simulations suggest that a
density constrast between the cold material and the ICM of $10^{4}$ is
a good description, and that a wind of velocity $\sim 6 \times
10^{7}\,{\rm cm\,s^{-1}}$ is required to produce the observed
filamentary structures. Furthermore, the kinematic zig-zag features
only seem to form after about $30$ Myrs, suggesting that the Perseus
filaments are approximately this old.

In each case, the total mass removed from the cold cloud was of the
order of $10^{9}\,M_\odot$. However, the majority of this material is
not in the filament, but is in the more amorphous and disrupted region
behind the cloud. The material around the cloud is likely to have a
low ionisation and is therefore not likely to be optically emitting.

\section{Momentum transfer from the ICM}

The kinematics of the filaments indicate that momentum transfer
between the cool and hot phases of the ICM occurs. If there is a large
enough quantity of cool gas, this could have important consequences
for the flow of the hot phase. Momentum $P$ is transferred from the
hot ICM phase to a cloud at a rate
\begin{equation}
\dot{P}_{\rm clump} = KA\rho_{\rm w}(v_{\rm
w}-\dot{x})^{2},
\end{equation}
If the radius of the cloud is $\sim 30$pc, and $(v_{\rm w}-\dot{x})
\sim 10^{7-8}{\rm cm\,s^{-1}}$ then $\dot{P}_{\rm clump} \sim
10^{29-30}{\rm erg\,cm^{-1}}$.  

A rough estimate for the rate at which momentum is transferred from
the external medium to the cloud's tail is
\begin{equation}
\dot{P}_{\rm tail} = m_{\rm tail}\ddot{x}_{\rm tail},
\end{equation}
where $m_{\rm tail} = A\rho_{\rm tail}l_{tail}$, $\rho_{\rm tail}$ is
the density in the tail, $\ddot{x}_{\rm tail}$ is the typical
acceleration in the tail, and $l_{\rm tail} = l_{\rm f}$ is the length
of the tail. Constant cross-sectional area, density, and acceleration
along the tail are assumed. The acceleration can be approximated by
$\ddot{x}_{\rm tail} = {(\dot{x} - u)^{2}}/2l_{\rm tail}$ where
$\dot{x}$ is the velocity of the clump, and $u$ is the maximum
velocity of the tail material. If the number density in the tail is
about 10 $\rm cm^{-3}$, the tail radius is about 50 pc, and the
magnitude of $|u-\dot{x}|$ is $\sim 10^{2}$ km $\rm s^{-1}$, then
\begin{equation}
\dot{P}_{\rm tail} \sim 10^{32}\bigg(\frac{n}{10\,{\rm
cm^{-3}}}\bigg)\bigg(\frac{R}{50\,{\rm
pc}}\bigg)^{2}\bigg(\frac{\dot{x}-u}{100\,{\rm km\,s^{-1}}}\bigg)^{2}{\rm
erg\,cm^{-1}}
\end{equation}
The corresponding rate at which systemic kinetic energy is transferred
to the tail is $\dot{E}_{\rm tail} \sim \dot{P}_{\rm tail}|(u-\dot{x})|$,
\begin{equation}
\dot{E}_{\rm tail} \sim 10^{39}\bigg(\frac{n}{10\,{\rm
cm^{-3}}}\bigg)\bigg(\frac{R}{50\,{\rm
pc}}\bigg)^{2}\bigg(\frac{\dot{x}-u}{100\,{\rm km\,s^{-1}}}\bigg)^{3}{\rm
erg\,s^{-1}}
\end{equation}
If the energy dissipated as thermal energy were comparable to this and
all of that thermal energy were radiated in the observed optical
lines, about 100 tails would account for the observed optical
luminosity of the NGC 1275 filaments.

The estimate above suggests that the tails are extremely massive $\sim
10^{8}\,{\rm M_\odot}$, though the mass of ionised material must be
much lower \citep[e.g.][]{hatch07}. This suggests that the filaments
may consist of molecular material surrounded by a skin of ionised
material.

We can compare the momentum transfer between the phases of the ICM,
with that developed by an AGN jet, which may in turn drive a wind from
the central galaxy. A jet injecting material at $1\, {\rm
M_{\odot}\,yr^{-1}}$ with a velocity of 0.1 times the speed of light
over a lifetime of 10 Myrs, has $\dot{P}\sim 10^{50} {\rm
g\,cm\,s^{-1}}$. This can be compared to the total momentum
transferred along the length of a filament by the wind $\sim
\dot{x}Al_{\rm tail}\rho_{\rm tail} \sim 10^{48}{\rm
g\,cm\,s^{-1}}$. This suggests that the momentum transfer to many
tails could be important for dissipating the energy injected by an AGN
in the ICM. In particular, since much of the cold material is located
near the centre of the cluster, if this dissipation mechanism is
significant, then it will increase the AGN heating efficiency in the
central regions of the cluster. Without this obstructing cold
material, more of the injected energy would be deposited at larger
radii, thus failing to heat the central regions where energy is most
needed.

To summarise this point it is worth reiterating that tails only seem
to form if there is a significant outflowing wind - they are formed by
the ambient wind dragging and accelerating material, transferring
energy and momentum to the cold material. As a result they are
probably a sign of activity in the central galaxy. This mechanism can
provide a way of coupling the wind to the cold material. Given that
the cold clouds are likely to proliferate near the cluster this means
that the energy will be dissipated more centrally than in the absence
of the cold material. As such, the cold material increases the
effective opacity of the ICM to the outflowing wind.

To estimate this effect, let us assume that energy is injected at a
rate $\dot{E}_{\rm in}$, the energy dissipation rate per tail is
$\dot{E}_{\rm tail}$, and there are $N$ tails. The change in power of
the wind, over a region in which there are $N$ clouds, will be $\Delta
\dot{E} = \dot{E}_{\rm in} - N\dot{E}_{\rm tail}$. Taking the
derivative with respect to radius, and assuming that $\dot{E}_{\rm
tail}$ is a constant we find,
\begin{equation} \label{eq:taildiss}
\frac{{\rm d}\dot{E}}{{\rm d}r} = - \frac{{\rm d}N}{{\rm
d}r}\dot{E}_{\rm tail},
\end{equation}
where ${\rm d}N/{\rm d}r$ is obtained from the spatial distribution of
the clouds,
\begin{equation}
\frac{{\rm d}N}{{\rm d}r} = 4 \pi r^{2}n(r)
\end{equation}
and $n(r)$ is the concentration of clouds per unit volume. If we
assume that the clouds are distributed similarly to the hot gas, in a
$\beta$-profile, then
\begin{equation}
n(r) = \frac{n_0}{(1 + (r/r_0)^2)^\beta},
\end{equation}
where $n_0$ is the central density and $r_0$ is a scale length.
Choosing $\beta = 1$ means that equation (\ref{eq:taildiss}) can be
integrated analytically to give,
\begin{equation}
\dot{E}(r) = \dot{E}_{\rm in} - N_{0}\dot{E}_{\rm
tail}\bigg[\frac{r}{r_0} - \arctan\bigg(\frac{r}{r_0}\bigg)\bigg], 
\end{equation}
where $N_0 = 4\pi n_0 r_{0}^{3}$. Therefore, the injected energy would
be dissipated over a typical length-scale
\begin{equation}
r_{\rm diss} \sim \frac{r_{0}\dot{E}_{\rm in}}{N_0\dot{E}_{\rm tail}}.
\end{equation}
Given that $r_0$ is probably a constant of the system, and
$\dot{E}_{\rm tail}$ is also probably fairly constant, the important
parameters are $\dot{E}_{\rm in}$ and $N_0$. For a moderate AGN
outburst, $\dot{E}_{\rm in} \sim 10^{43}\,{\rm erg\,s^{-1}}$, with
$\dot{E}_{\rm tail} \sim 10^{40}\,{\rm erg\,s^{-1}}$ and $N_0 \sim
1000$ (corresponding to a total cold mass of $10^{11}\,M_{\odot}$),
the dissipation length is of the same order as the scale-height of the
cloud distribution. (This may be comparable to the scale-height of the
ICM, but that is not clear. It is also possible that the clouds are
more centrally concentrated than the hot ICM, but again, this is
speculation.)  For more powerful outbursts, a significant quantity of
energy may be trapped, but the majority will still escape outside
$r_0$.

It is worth noting that if there has been significant recent cooling
(i.e. a deficit of heating) we may expect $N_0$ to be larger, for a
given system. For a given energy injection rate, the energy will
necessarily be dissipated closer to the centre than in an otherwise
identical system where $N_0$ is lower. As a result, the presence of
cold gas may provide an additional feedback process that helps to trap
energy in the central regions. This effect will be greater if there
has been a recent deficit of heating, with the trapping effect being
reduced in the absence of cold material.

This short derivation is, of course, a drastic simplification of real
life. For example, in reality, the power dissipated in a tail is a
reflection of the power of the incident wind, which will undoubtedly
be a function of radius. However, it highlights the possibility that
the production of filaments may be able to trap a fraction the energy
injected by a central AGN.

\section{General expectations for optical emission-line filaments in galaxy clusters}

The current study is centred on the Perseus cluster and concerns the
interaction of material injected from one cloud into a hot flow
unaffected by other clouds. A global consideration of the effects of a
distribution of clouds on the hot flow out to $\sim$10 kpc would be
required to investigate the problem more completely, and we plan to
investigate such problems. However, a full answer would require the
developement of a detailed theory for the spatial and mass
distributions of the clouds; given the state of the theory of our own
interstellar medium this task would be very difficult.

The nature and number of filaments will depend on: a) the spatial,
mass and velocity distributions of cold clouds at the onset of AGN
activity; b) the rate at which the AGN deposits mechanical and thermal
energy into its outflow and the asymmetry of these AGN outflows; c)
the pressure, velocity, and density distributions of the hot
intracluster gas at the onset of AGN activity; d) the gravitational
potential of the galaxy. Below we consider, in sequence, how each of
these could affect the filamentary structure.

a) If all other factors were the same, the total number of filaments
and their combined optical luminosity would be roughly proportional to
the number of clouds that could survive interactions with an outflow
long enough for filaments to evolve. If the distribution of cloud
masses were identical in all clusters, the systems with the most cold
gas would exhibit the most luminous optical emission from
filaments. This is consistent with the findings of \cite{edge01}. The
clouds which survive for longest are likely to be the most massive.

b) The number of massive clouds will depend on AGN activity. If the
clouds are formed in a steady fountain flow, the rate at which
material turns into clouds would be proportional to the outflow rate
of hot material as measured near its the source. Roughly speaking,
this mass outflow rate may be proportional to the mechanical
luminosity of the AGN wind. In addition, the destruction rate of the
clouds may also be proportional to the outflow rate. For example, a
higher outflow rate could lead to a more rapid destruction rate of the
clouds. Thus, it is possible that the number of clouds producing
filaments would be fairly insensitive to the AGN mechanical
luminosity. However, it would be cavalier to assume the locations at
which clouds form, and the initial formation of the clouds, to be
independent of the outflow rate.

c) The properties of the ICM could affect the initial mass function of
the clouds. If significant, thermal conduction might establish the
minimum lengthscale on which thermal instability grows. In hotter
clusters, the minimum cloud mass would be higher, leading to a greater
fraction of the clouds being formed with high masses. This would tend
to increase the number of long filaments.

High ICM densities will lead to more rapid cooling and possibly a
higher formation rate of clouds, and hence more filaments. However,
given that AGN feedback is likely to be important, high cooling rates
may cause strong AGN activity and thus reduce the cloud formation
rate. The exact balance between heating and cooling seems to vary from
cluster to cluster.

Clouds in the ICM may be so plentiful in the centre of a cluster that
a bubble blown by the AGN becomes trapped near the cluster centre,
preventing the formation of long filament. So, high ICM density need
not necessarily result in the production of many filamentary tails.

d) The gravity of the central galaxy may play a role in limiting the
size of the bubble mentioned in c). For example, the greater buoyancy
may cause a bubble to detach sooner. Strong gravity may prevent an
extended fountain or region of outflows. Even if it does not, it may
limit the lifetimes of clouds as it would cause them to fall more
rapidly.

In summary, the question about scaling concerns very complex issues. A
very long concentrated program would be required to address them
properly, and it is unclear how reliable the results of such a program
would be.

\section{Summary}

This work is intended mainly as a preliminary study into the processes
that shape filamentary structures in the ICM, and not an exhaustive
study. It is likely that there are many other processes such as
magnetic fields, thermal conduction and viscosity that all play a role
in the dynamics that produce the filamentary kinematics and
morphology. However, we have concentrated on the basic hydrodynamic
system and have found several interesting mechanisms that may go some
way to explaining the properties of the observed filaments.

The first ensemble of simulations show that the long, relatively
straight filaments at the centre of the Perseus cluster may be formed
by the interaction of a wind with clumps of cold material. Flows of
low Mach number winds generate structure in the filaments which may be
comparable with observations. These results also suggest that the
amplitude of the velocity fluctuations along the length of the
filament grow with increasing density contrast between the cold
material and the ambient wind. The morphology of the observed
filaments suggests that the density contrast is large ($\sim
10^{4}$). Interestingly, the fluctuations evolve with time providing a
possible diagnostic tool for estimating the ages of filaments from
observations. in this set of simulations, the best comparison with
observations suggests filament ages of $\sim 40$ Myrs in the Perseus
cluster.

In a more realistic environment with spatially varying gravity, based
on the Perseus cluster, filaments only form if the wind velocity is
sufficiently large. If there is no significant wind, then the optical
emission may be more amorphous, or more likely, filamentary on much
shorter length scales. The morphology and kinematics suggests that a
density contrast of $\eta \sim 10^{3}$ may be more compatible with the
observations, while the best agreement in the kinematic data occurs
for $\eta \sim 10^{4}$. As a compromise it is possible that the real
conditions lay somewhere between these two extremes. Alternatively,
other physical processes could be important. It should also be noted
that these were 2-d simulations, and a full 3-d study would really be
required to find the model conditions that best match
reality. Regardless of this, the results do point to a relatively
narrow range of parameters providing possible information about the
state of the ICM, and the age of the filaments. These results suggest
that the cold material is of the order of $10^{4}$ times denser than
the ambient ICM, there is an outflowing wind of velocity of $\sim
10^{3}\,{\rm km\,s^{-1}}$, and that the filaments are approximately a
few 10's of Myrs old.

The masses, and densities, of the filaments created by these
simulations seem to be low compared with values presented in the
previous section. The simulations produce filaments of
$10^{6}-10^{7}\,{\rm M_\odot}$, while the crude estimates above
suggested that $10^{8}\,{\rm M_\odot}$ filaments might be more
likely. It should be noted though, that the quantity of cold gas, in
the simulated tail, is more than capable of producing the observed
optical luminosity. The simulations also show a lot of cold, dense gas
in the vicinity of the cloud itself.

We also highlight the possibility that momentum transfer between the
ICM and cold clouds could be a dynamically important process in galaxy
clusters. This process could effectively couple the energy and
momentum of an outflowing wind to the cold material, thereby
dissipating some of the energy injected by a central AGN. The presence
of cold gas may therefore provide an additional feedback mechanism
that traps energy in the central regions of clusters where it is most
needed.

\section{Acknowledgements}

We acknowledge Professor John E. Dyson and Dr Nina Hatch for many
useful discussions and the anonymous referee for helpful comments.

\bibliography{database} \bibliographystyle{mn2e}

\label{lastpage}

\end{document}